\documentclass[doublecol]{epl2} 
\usepackage{graphics}
\usepackage{amssymb}
\usepackage{subfigure}
\usepackage{amsmath}
\usepackage{color}
\usepackage{hyperref}
\usepackage[normalem]{ulem}

\newcommand{\fig}[1]{Fig.~\ref{#1}}

\title{Differences in the scaling laws of canonical and microcanonical coarsening dynamics for long-range interacting systems}

\author{F. Staniscia\inst{1}\footnote{\email{fabiostaniscia@gmail.com}}, R. Bachelard\inst{2}, T. Dauxois\inst{3}, G. De Ninno\inst{1,4}}
\institute{\inst{1} Laboratory of Quantum Optics, University of Nova Gorica, 5001 Nova Gorica, Slovenia\\ 
\inst{2} Universidade Federal de S\~{a}o
Carlos, Rod. Washington Luis, km 235, S/n - Jardim Guanabara, S\~{a}o
Carlos - SP, 13565-905, Brazil\\ 
\inst{3}Universit\'e de Lyon,  ENS de Lyon, UCBL, CNRS, Laboratoire de Physique, Lyon, France\\
\inst{4} Sincrotrone Trieste, S.S. 14 km 163.5, Basovizza (Ts), Italy 
}
\shortauthor{F. Staniscia \etal}

  
\date{\today} 

\abstract{We investigate the effects of Hamiltonian and Langevin microscopic dynamics on the growth laws of domains in coarsening. Using a one-dimensional class of generalized $\phi^4$ models with power-law decaying interactions, we show that the two dynamics exhibit scaling regimes characterized by different scaling laws for the coarsening dynamics. For Langevin dynamics, it 
concurs with the exponent of defect dynamics, while Hamiltonian dynamics reveals new scaling laws with distinct early-time and a late-time regimes. This new behaviour can be understood as an effect of energy conservation, which induces a coupling between the dynamics of the local temperature field and of the order parameter.}

\begin{document}

\maketitle

\section{Introduction}\label{sec:intro}
If an Ising model is quenched from a high temperature disordered equilibrium state to temperatures below the critical one, 
coarsening takes place~\cite{Bray2002}. Coarsening manifests itself by the emergence of ordered ferromagnetic domains, 
and the subsequent scale-free growth of the larger domains at the expense of the smaller ones~\cite{Puri2009}. 
This phenomenon has been mostly studied for models with nearest neighbour interactions~\cite{Bray2002} in absence or
presence of disorder~\cite{Sicilia2008}.
The theory is based on the hypothesis that two point spatial correlations are time invariant, provided that the distances are renormalized with a 
time-dependent length $L(t)$ which usually, at leading order, scales as $t^{1/z}$. This scaling hypothesis has been rigorously demonstrated 
for one-dimensional models~\cite{Amar1990,Bray1990} and for the Ginzburg-Landau model in the limit of infinite components of the order parameter~\cite{Coniglio1989}. However, simulations and experiments indicate its wider applicability~\cite{Bray2002,Puri2009,Noda84}. Most of the studies were carried out for systems coupled to a thermal bath. Nonetheless, some authors~\cite{Zheng,ChateReply,Zheng_rep,Chate2} have performed simulations of
the two-dimensional isolated $\phi^4$ model with nearest-neighbour couplings, verifying the scaling hypothesis but without finding agreement on whether the scaling exponents are the same as when coupled to a bath.

Regarding systems with long-range couplings, coarsening has been also analysed theoretically and numerically for two simple one-dimensional lattice models (Ising and $\phi^4$) with long-range 
couplings~\cite{Cardy1993,Bray1994b}. In these models, the coupling decays with the lattice distance as $(r_{i,j})^{-(1+\sigma)}$ at large distances 
$r_{i,j}$ between pairs of lattice sites ($i,j$). Coarsening has been found at finite temperature if $0<\sigma \le 1$ and at zero temperature for
$\sigma>1$. Using an effective model for the time evolution of sharp domains boundaries, these authors find that $L(t) \sim t^{1/(1 + \sigma)}$. 

To our knowledge the question whether the dynamical exponent $z$ is the same for systems coupled to a thermal bath, whose equilibrium corresponds to the canonical ensemble, and to isolated systems, whose equilibrium corresponds to the microcanonical ensemble, for systems with long-range
interactions has not been addressed before. From the point of view of equilibrium thermodynamics, the range of parameter values $0 < \sigma \le 1$ of the models, is characterized by the equivalence of microcanonical and canonical ensembles~\cite{Mukamel,reviewCDR} (although a subtle dependence on the ensemble may exists also at equilibrium~\cite{Kastner2000}), but this equivalence is not guaranteed for dynamical phenomena, for which reason we extend the notion of canonical and microcanonical ensemble to the out-of-equilibrium regime.

In this letter, we want to verify the scaling hypothesis for both the Langevin (canonical) and the Hamiltonian (microcanonical) microscopic dynamics.  
More importantly, we aim at checking whether the dynamical scaling exponent $z$ is the same in both dynamics and agrees or not with that found 
in~\cite{Cardy1993,Bray1994b}, i.e. $z=1+ \sigma$. To this purpose we consider a $\phi^4$ model~\cite{Desai78,StefanoThierry} with long-range 
couplings, which displays coarsening in the relevant range of values, $0 \leq \sigma \leq 1$. 

We will show that, by a careful numerical analysis of the spatial correlation function for different values of $\sigma$ it is possible to validate the scaling hypothesis in both the canonical and the microcanonical ensemble. However, the dynamical scaling exponent is
found to be sharply different in the two ensembles: we obtain the law $z_{c}=1+\sigma$ in the canonical ensemble, in agreement with Refs~\cite{Cardy1993,Bray1994b}, while, in the microcanonical we get $z_{\mu}^{early}=2\sigma$ (at early times) and $z_{\mu}^{late}=2$ (at late times). Moreover, the two types of dynamics differ by additional dynamical features as, for instance, the scaling regimes appears on different time scales in the two ensembles and the Hamiltonian dynamics shows transient
oscillations of $L(t)$ just before the scaling regime sets in, which is the signature of a collective phenomenon. Additionally the structure factor shows a power-law tail in the Langevin case, in agreement with Porod's law, which does not appear in the Hamiltonian case.  

\begin{figure}
  \includegraphics[width=\linewidth]{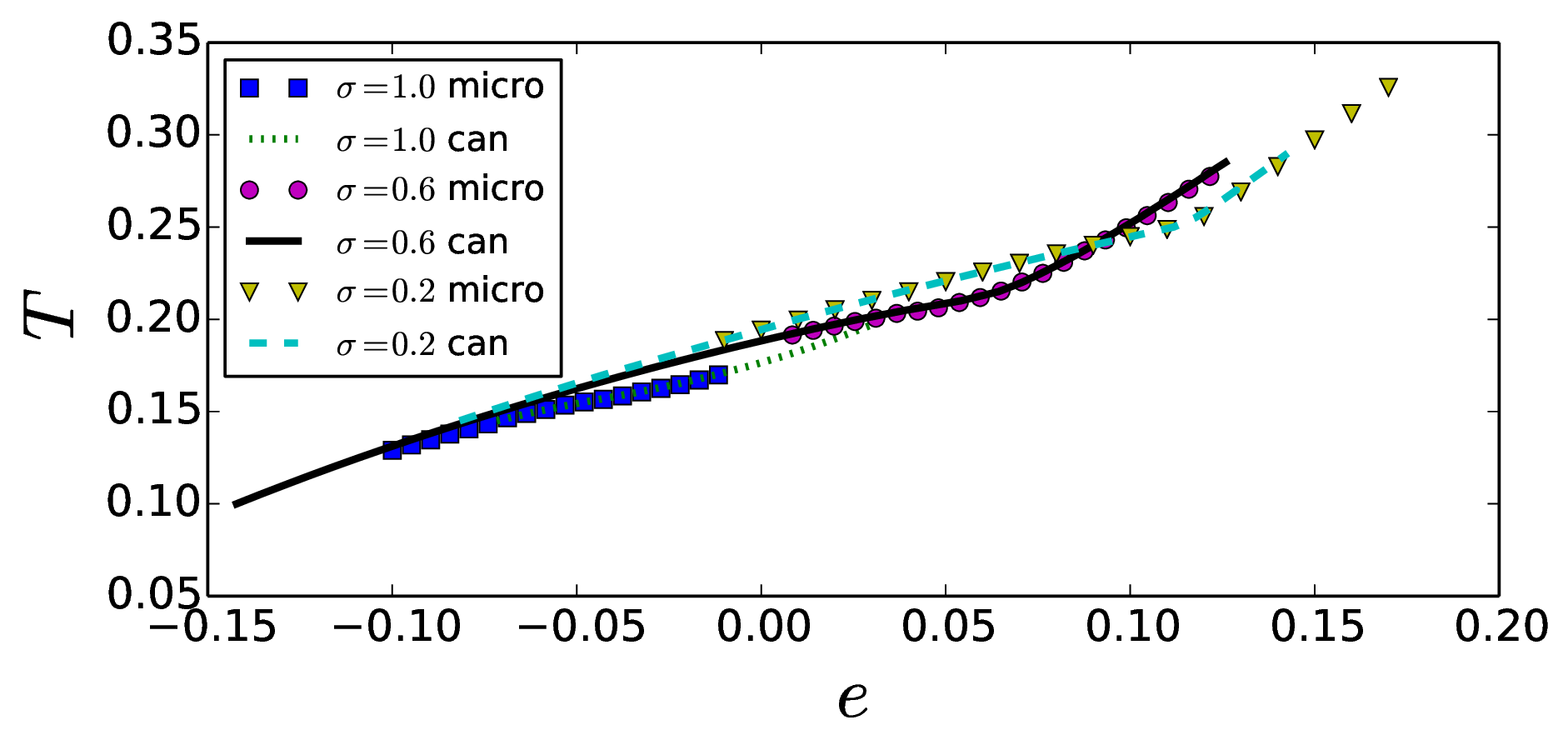}
  \includegraphics[width=\linewidth]{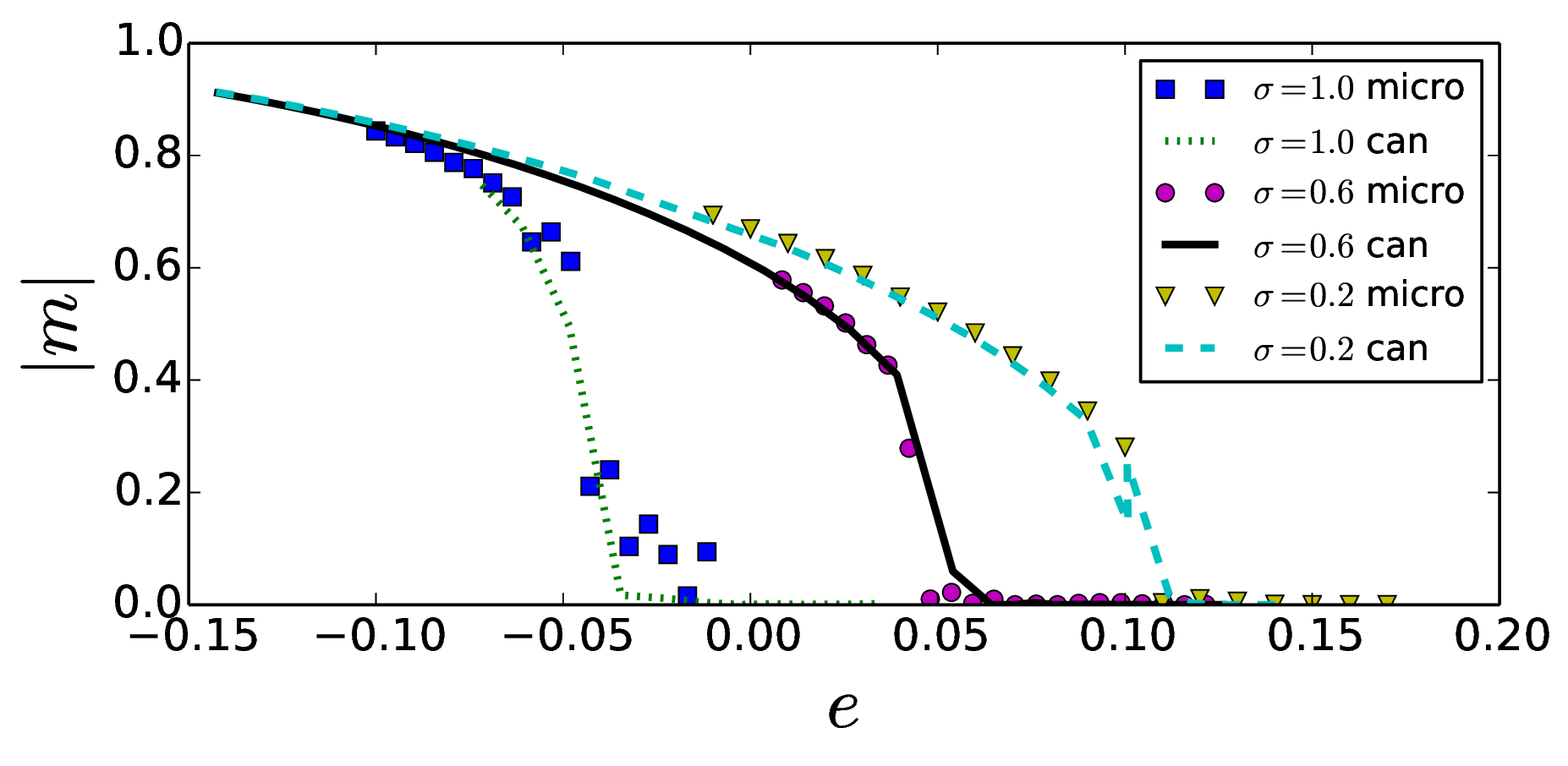}
  \caption{\label{f:equilibrium} Top: temperature $T$ vs. energy per particle $e$. Bottom: magnetization $m$ vs. energy per particle $e$. The range of parameters simulated has been chosen to be around the transition point for each value of $\sigma$. In all cases $N=8192$. }
\end{figure}

\section{The $\phi^4$ model} \label{model}
We consider a one dimensional periodic lattice of $N$ sites. To each lattice site $i$, we attach a scalar variable
$q_i$, with $i=1,\dots,N$. The potential energy is defined as
\begin{equation}
\label{e:ham}
  U = \sum_{i=1}^{N}\left( \frac{q_i^4-q_i^2}{4} \right)
     - \frac{1}{4 \tilde{N}} \sum_{i\neq j=1..N} \frac{q_i \, q_j}{r_{ij}^{1 + \sigma}}\,,
\end{equation}
where $r_{ij}=\mbox{min}(|i-j|,N-|i-j|)$ is the closest distance on the periodic lattice between sites $i$ and $j$ and 
$\tilde{N}=\sum_j r_{ij}^{-(1+\sigma)}$ is a normalization factor which makes the energy extensive in $N$ even when $\sigma\le 0$. The scalar variable
$q_i$ can be viewed as representing a local magnetization. This magnetization feels the action of the on-site potential 
$(q^4-q^2)/4$, which favours the two magnetization values $q=\pm 1/\sqrt{2}$, and the effect of the long-range ferromagnetic coupling.
The order parameter of the model is the total magnetization
\begin{equation}
  m=\frac1N\sum_{i=1}^N q_i.
\end{equation}
A model with the same symmetries and the same interactions as~\eqref{e:ham}, the Ising model with long-range couplings, has been originally studied by Ruelle~\cite{Ruelle68} and Dyson~\cite{Dyson69a} in the
canonical ensemble. In the range $0 \leq \sigma <1$, this model undergoes a second order phase transition~\cite{Dyson69a} separating 
a ferromagnetic phase ($m \neq 0$) at low temperatures from a paramagnetic phase ($m=0$) at high temperatures. For $\sigma>1$, 
the system is disordered at all energies~\cite{Ruelle68,Dyson69b}. The case $\sigma =1$ is peculiar, since 
it shows a Kosterlitz-Thouless phase transition with a discontinuous jump in the magnetization~\cite{Thouless69,Aizenman1988,Luijten01,Theodorakopoulos2006185,Bar13}. 
At equilibrium, the $\phi^4$ model~\eqref{e:ham} exhibits the same qualitative features of the Ruelle-Dyson model in both the microcanonical and canonical ensembles.

One can study the dynamics of model \eqref{e:ham} in the canonical ensemble by coupling each site to a heat reservoir at constant 
temperature $T_{can}$. This can be done by considering the over-damped Langevin equations, 
\begin{equation}
\label{e:eomLang}
 \gamma \, \dot{q}_i + \frac{\partial U}{\partial q_i} = \eta(t) \quad \quad i=1,\ldots,N,
\end{equation}
where $U$ is given by \eqref{e:ham} and $\eta(t)$ is a zero average $\delta$-correlated Gaussian noise:
\begin{equation}
  \overline{ \eta(t) \eta(t^{\prime}) } = 2 \gamma T_{can}\, \delta(t-t^{\prime})\,,
\end{equation}
where the bar denotes averaging over noise. 

Alternatively, by adding kinetic energy $K$ to the potential energy $U$ defined in~\eqref{e:ham}, one obtains the Hamiltonian
\begin{equation}
\label{e:Hamiltonian}
H=K+U=\sum_{i=1}^N \frac{p_i^2}{2}+U,
\end{equation}
where $p_i$ is the momentum conjugate to $q_i$. Hamiltonian~\eqref{e:Hamiltonian} defines the dynamics in the microcanonical ensemble.

From the numerical point of view, the time consuming part of the algorithm lies in the calculation of the force $-\partial U/\partial q_i$ acting on site $i$, because of the all-to-all coupling. For periodic boundary conditions this can be efficiently done by using the Fourier representation of the coupling matrix $1/r_{ij}^{1+\sigma}$, as discussed in the Appendix of ref.~\cite{Potters12}. In this way, one obtains an algorithm that scales with the number of sites as $N \ln N$. The Langevin dynamics is integrated using a second-order algorithm~\cite{Honeycutt1992} while, for the Hamiltonian dynamics we have implemented a symplectic fourth-order algorithm~\cite{MclagAtela}. We have also tested the results against those obtained using other algorithms.

First of all, we have checked whether these algorithms reproduce the equilibrium features of the model in both the canonical and the microcanonical ensemble. In Fig.~\ref{f:equilibrium}, we show the caloric curve and the magnetization vs the energy per particle $e$ for $\sigma=0.2$, $\sigma=0.6$ and $\sigma=1.0$. The value of energy in the canonical ensemble is obtained from the temperature, using the caloric curve. The superposition of caloric curves can be considered as a convincing numerical evidence of ensemble equivalence for this model at equilibrium. 
The transition energy/temperature decreases as $\sigma$ is increased above zero until $\sigma=1$.
For $\sigma=0.2$ and $\sigma=0.6$, the system shows continuous phase transitions respectively at the energy per particle $e_c \simeq 0.120$ and $e_c \simeq 0.061$, corresponding to the temperatures $T_c \simeq 0.253$ and $T_c \simeq 0.21$ in the canonical ensemble. These values have to be compared with the theoretically known values, $e_c=0.132$, $T_c=0.264$, of the mean-field case $\sigma\le0$~\cite{StefanoThierry}. 
At $\sigma=1$, it is possible to see the jump in the magnetization at $e \simeq - 0.45$ and $T \simeq 0.16$.
Above $\sigma=1$, there is no numerical signatures of phase transition at finite temperature, in analogy with what is known for the Ruelle-Dyson Ising model. One can conclude that the $\phi^4$ model with long-range interactions~\eqref{e:ham} displays a very similar behaviour to the Ruelle-Dyson model, for what concerns equilibrium properties.

\section{Scaling of the two point correlation function}
\label{s:g}
We have performed coarsening numerical experiments using both canonical dynamics~\eqref{e:eomLang} for model~\eqref{e:ham} and microcanonical
dynamics, derived from the Hamiltonian~\eqref{e:Hamiltonian}. We have considered quenches from the disordered phase to
a finite temperature/energy below the critical one and energy above the ground state, using the curves in \fig{f:equilibrium} and others computed for various values of $\sigma$ to determine the parameters of the initial distribution which assured a quenched initial condition.

\begin{figure*}[htb]
  \includegraphics[width=0.33\linewidth]{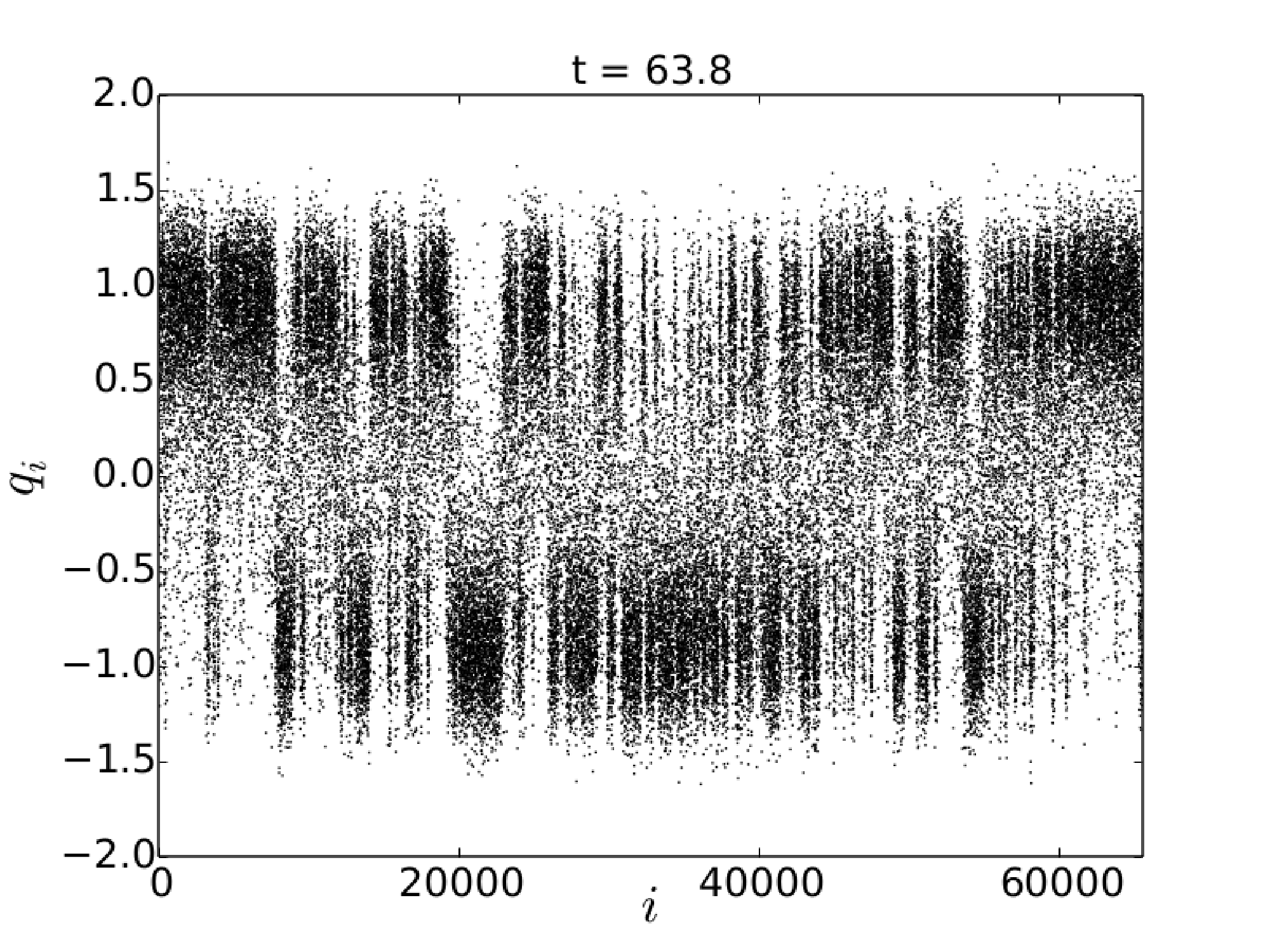}%{Fig2a}
  \includegraphics[width=0.33\linewidth]{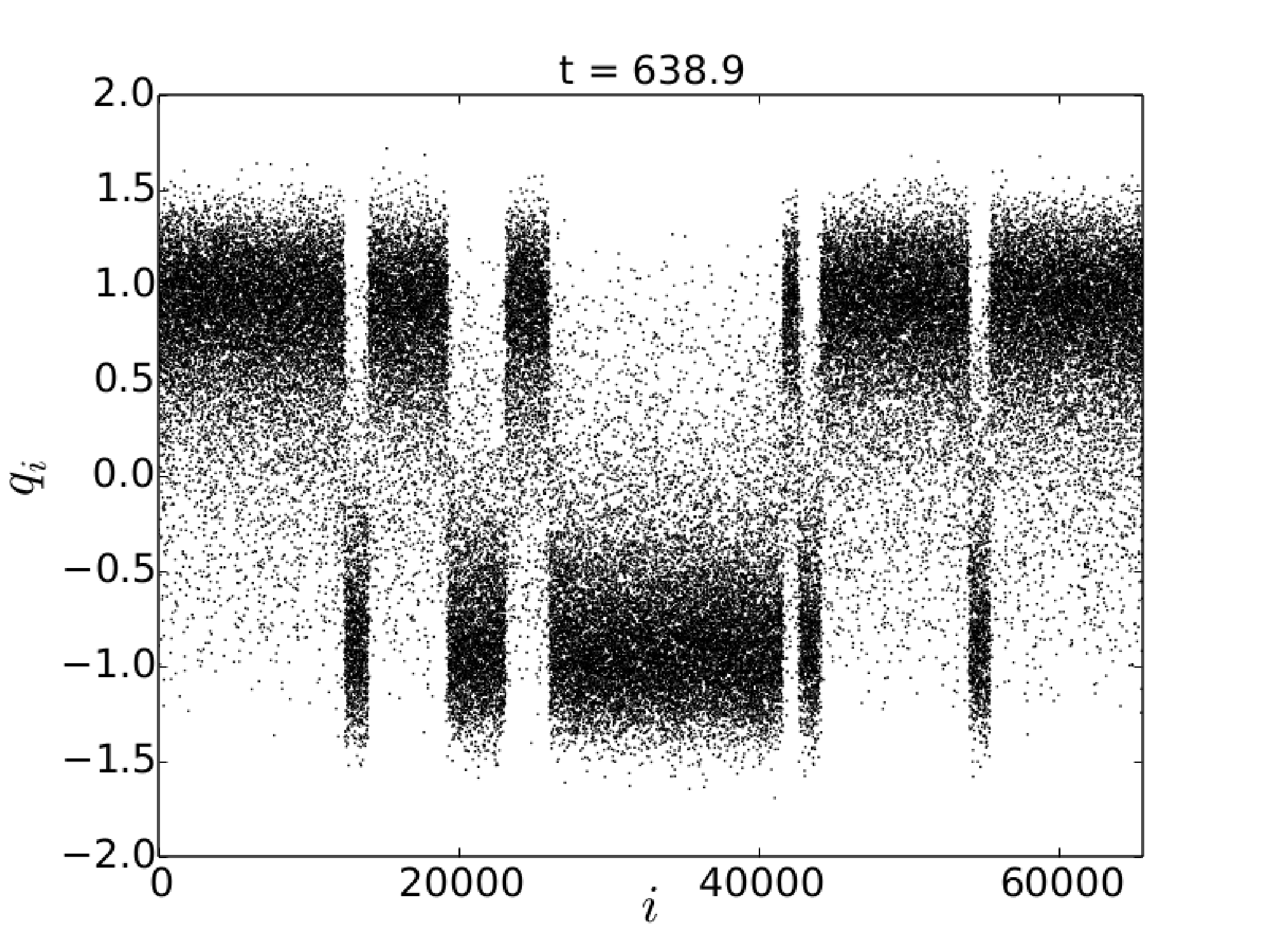}%{Fig2b}
  \includegraphics[width=0.33\linewidth]{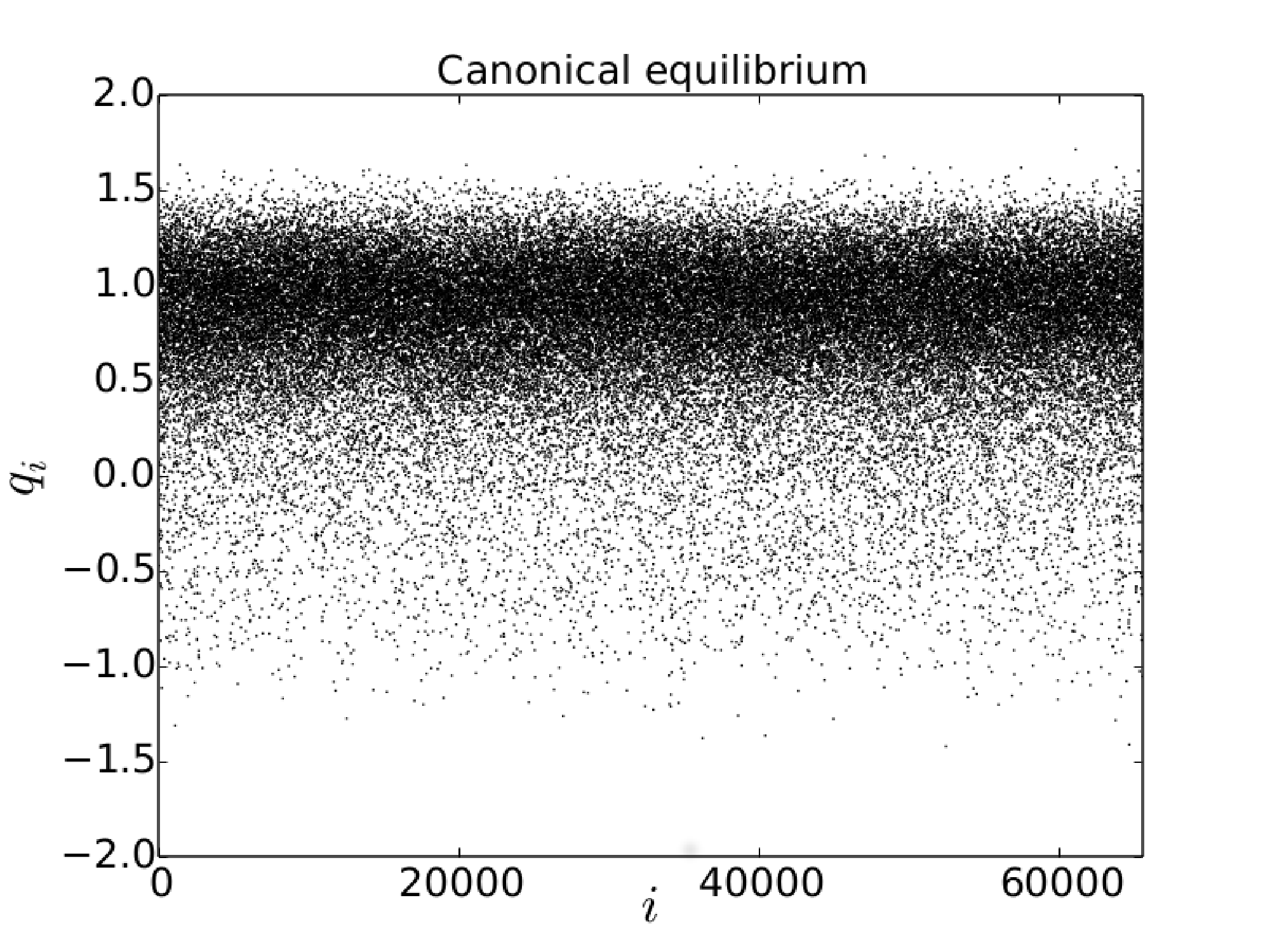}\\%{Fig2c}
  \includegraphics[width=0.33\linewidth]{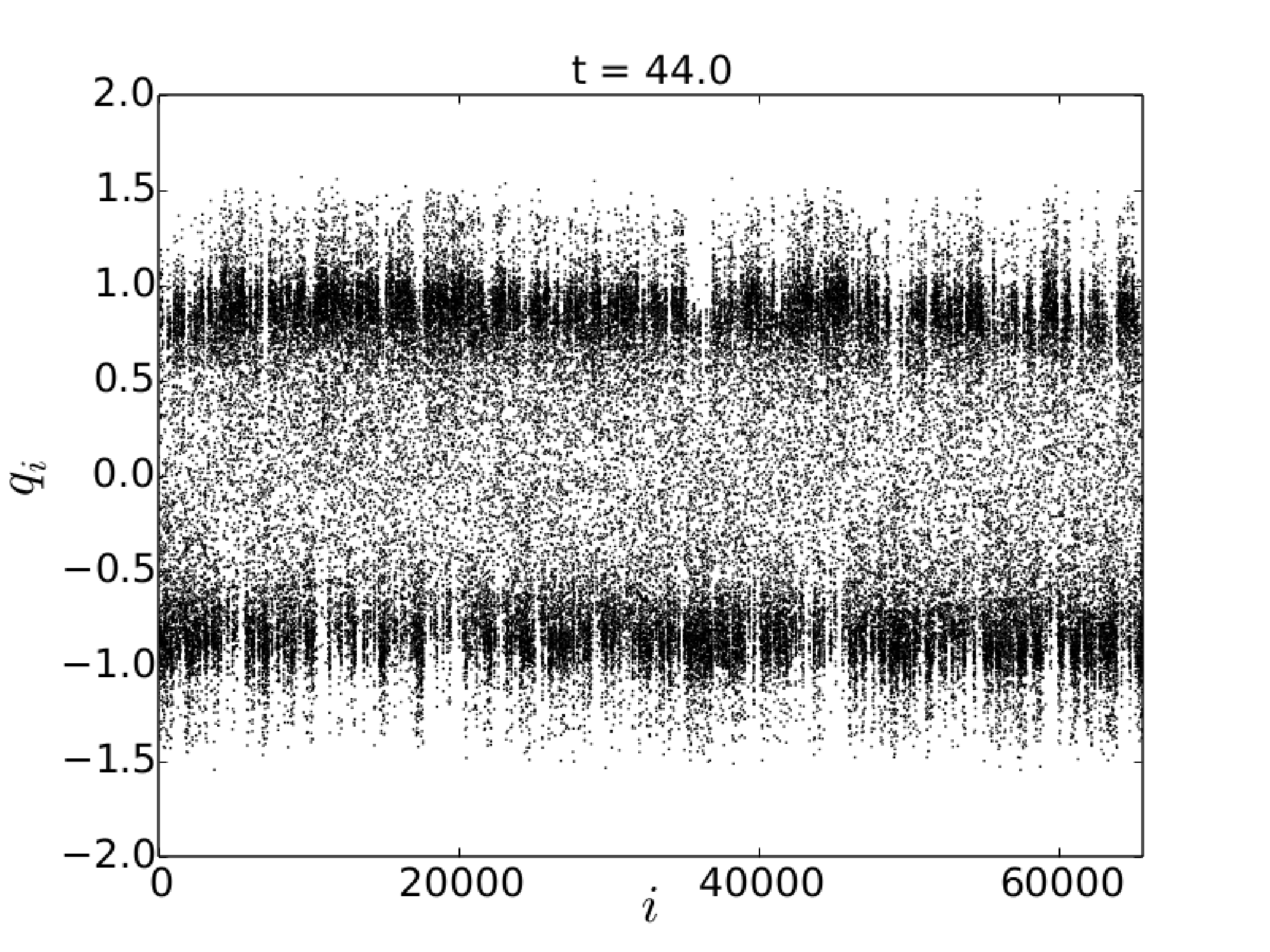}%{Fig2d}
  \includegraphics[width=0.33\linewidth]{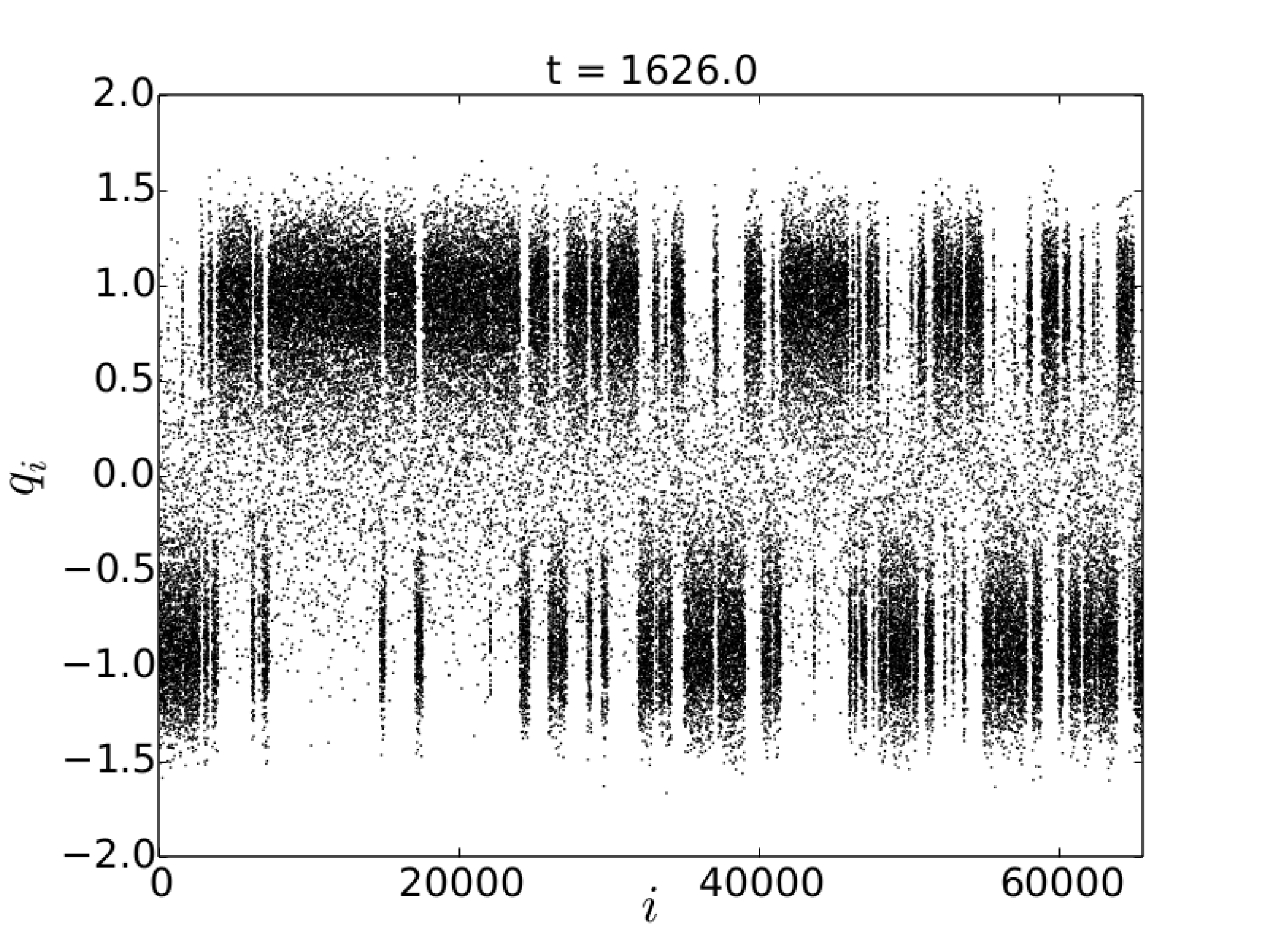}%{Fig2e}
  \includegraphics[width=0.33\linewidth]{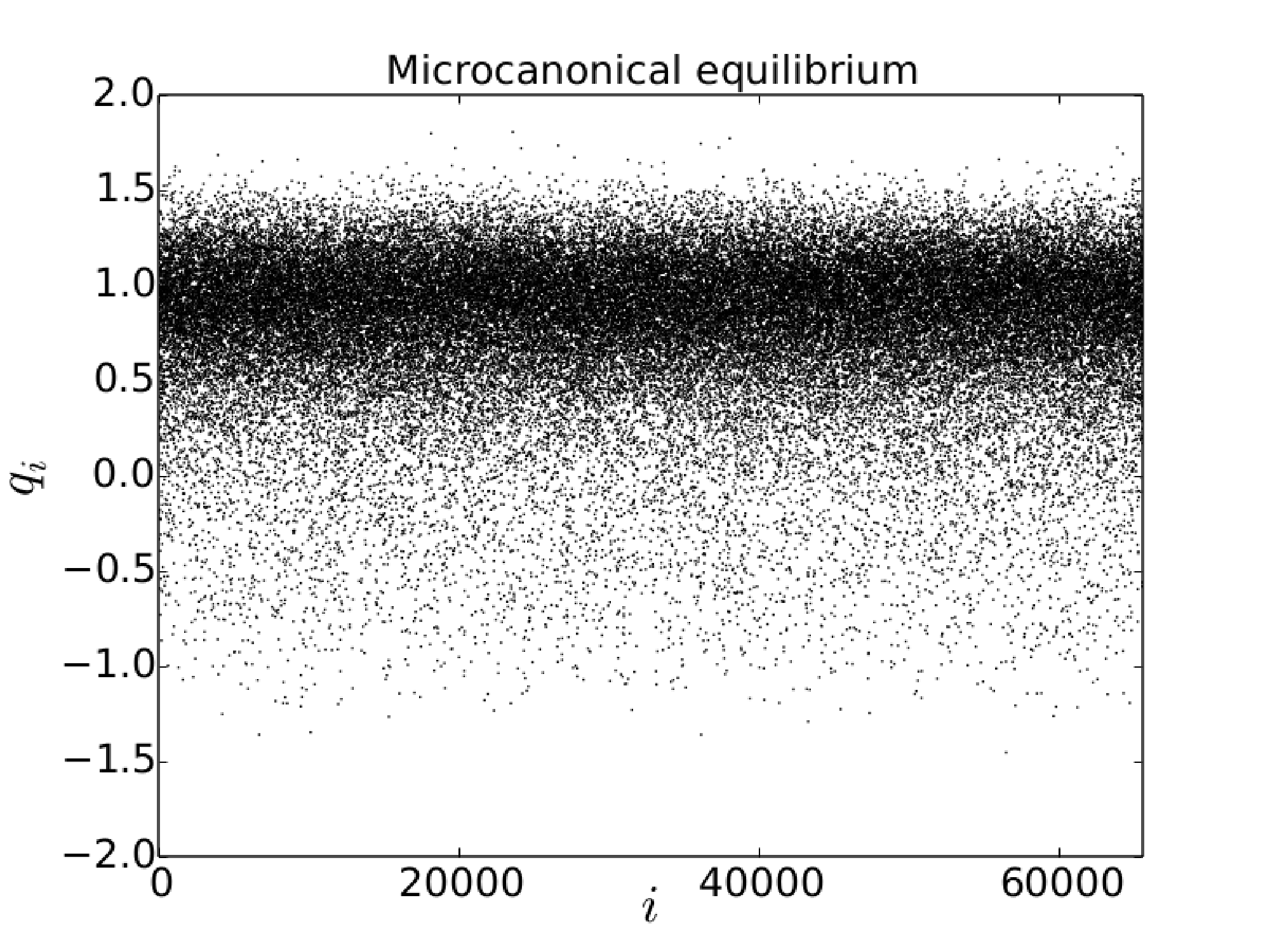}%{Fig2f}
  \caption{\label{f:domains}Quenched below a phase transition, Langevin (first row) and Hamiltonian (second row) evolutions undergo a coarsening process (left and middle columns) before eventually reaching a broken phase equilibrium (right column). The dots represent the vale of $q_i$ for each particle, with $i$ being the position on the lattice. Here $N=65536$, $\sigma=0.4$ in both cases, $T = 0.16$ and $\gamma=1$ for the Langevin case and $e = -0.06$ for the Hamiltonian case.} 
\end{figure*}

The initial distribution of the positions has been taken, in both ensembles, as uniform in a region symmetric around the $q=0$ axis, and zero elsewhere. The region is either connected and centred around the maximum of the potential at $q = 0$, or disconnected and formed by two equal parts centred around the two minima of the potential. Their area are chosen in order to have the energy desired.  For the Hamiltonian case, we chose a distribution of momenta uniform in a connected region  symmetric around the $p=0$ axis and zero elsewhere. The reason for this choice is that in the microcanonical ensemble it is not possible, because of energy conservation, to set initial conditions corresponding to the equilibrium configuration above the critical point, if we want the system to be below it, as is usually done for quenches in the canonical ensemble.
As an example, Fig.~\ref{f:domains} presents the typical evolution of the system for the Hamiltonian and Langevin cases, where the appearance of domains with different local magnetization can be observed. Their average size grows in time until one of them reaches the system's size. The last configuration corresponds to the equilibrium state. 

As expected, in the low temperature/energy region there is thus formation of domains as the system relaxes from a disordered to an ordered configuration, a regime during which we can extract the scaling law for their growth.

\begin{figure}[t]
  \includegraphics[width=\linewidth]{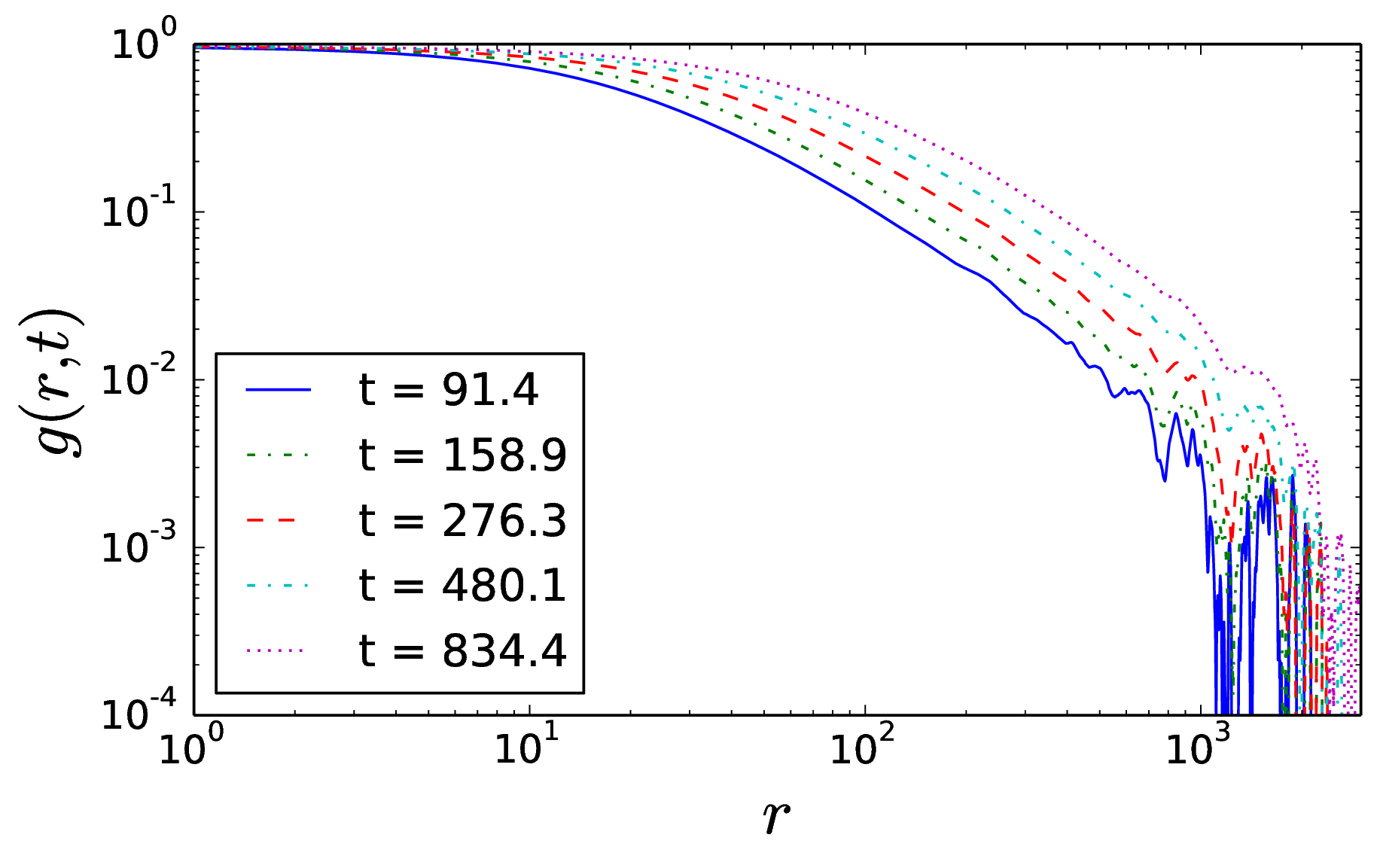}\\
  \includegraphics[width=\linewidth]{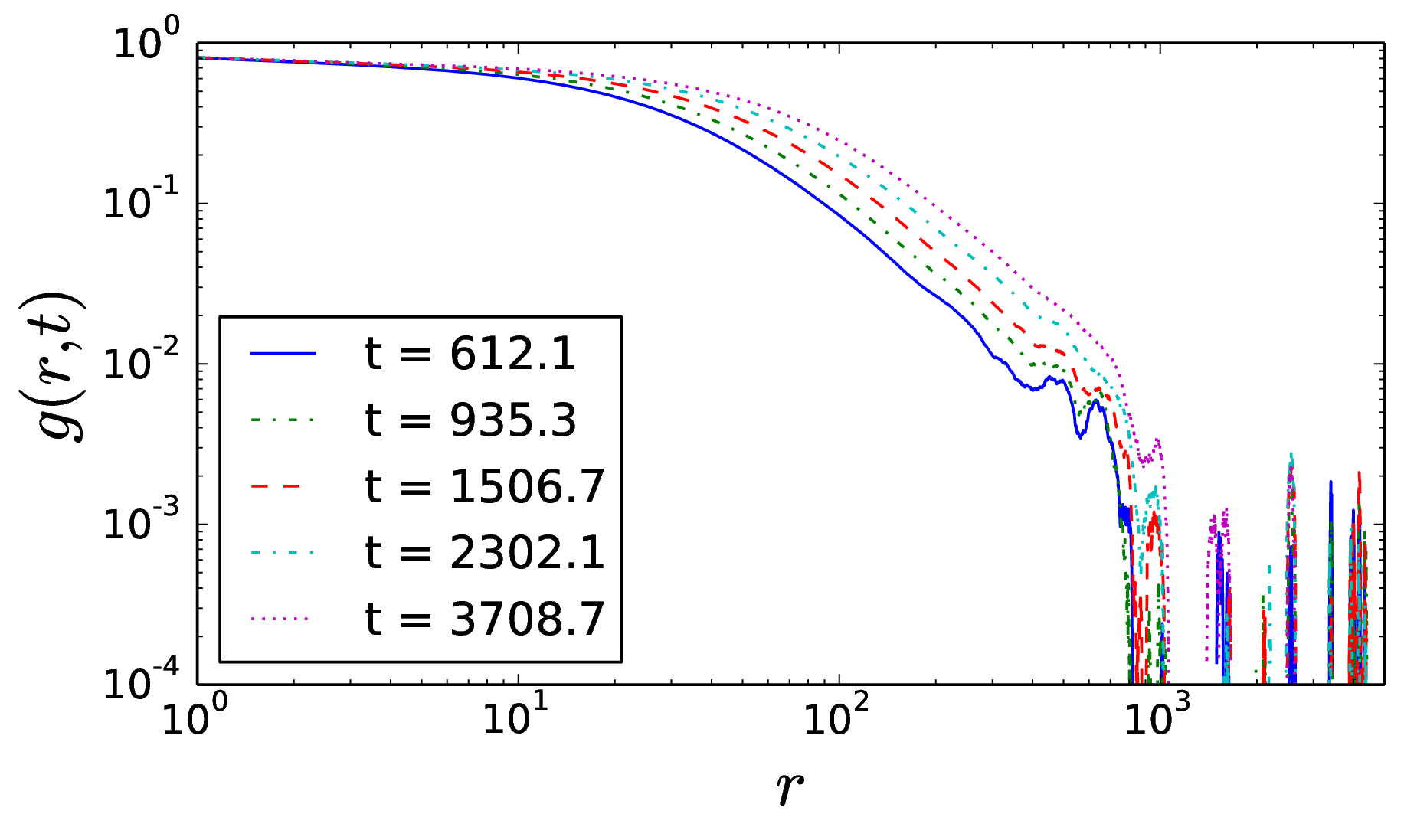}
  \caption{\label{f:g} Langevin (Top) and Hamiltonian (Bottom) correlation functions $g(r,t)$ at different times during the scaling regime.  Results shown are averaged over $400$ runs for systems with $N=65536$ and  $\sigma= 0.6$. Here $T = 0.05$ and $\gamma=1$ for the Langevin case and $e = -0.06$ for the Hamiltonian case.}
\end{figure}

To do so, we use the two-point correlation function $c(r,t)$, which is defined by:
\begin{equation}\label{e:g}
  c(r,t) = \langle q_i(t) q_j(t)\rangle_{r_{i,j}=r}  \,,
\end{equation}
where $\langle\cdot\rangle_f$ defines an average over the lattice subject to the constraint $f$. Several snapshots of the rescaled correlation function $g(r,t)=c(r,t)/c(0,t)$, for various $t$ are shown in~\fig{f:g}.

We are investigating the scaling hypothesis, which addresses the universality of the two-point correlation function:
\begin{equation}\label{e:hyp}
  g(r,t)\approx\tilde{g}(\tilde{r}(t))\, , \quad \mbox{for} \quad t_{\mathrm{transient}} < t < t_{\mathrm{cutoff}},
\end{equation}
where the scaled distance $\tilde{r}$ is defined as
\begin{equation}\label{e:sr}
  \tilde{r}(t)=r/L(t) \, .
\end{equation}
Let us define the scaling factor $L(t)$ as the distance at which the correlation function reaches some given fraction $1/n$ of its peak value:
\begin{equation}\label{e:L}
  L(t)>0~:~g(L(t),t)=g(0,t)/n \, .
\end{equation}
We chose the values of $n$ which give the best fit in the two ensembles, but we checked that different choices lead to the same conclusions. 

\begin{figure}
  \includegraphics[width=\linewidth]{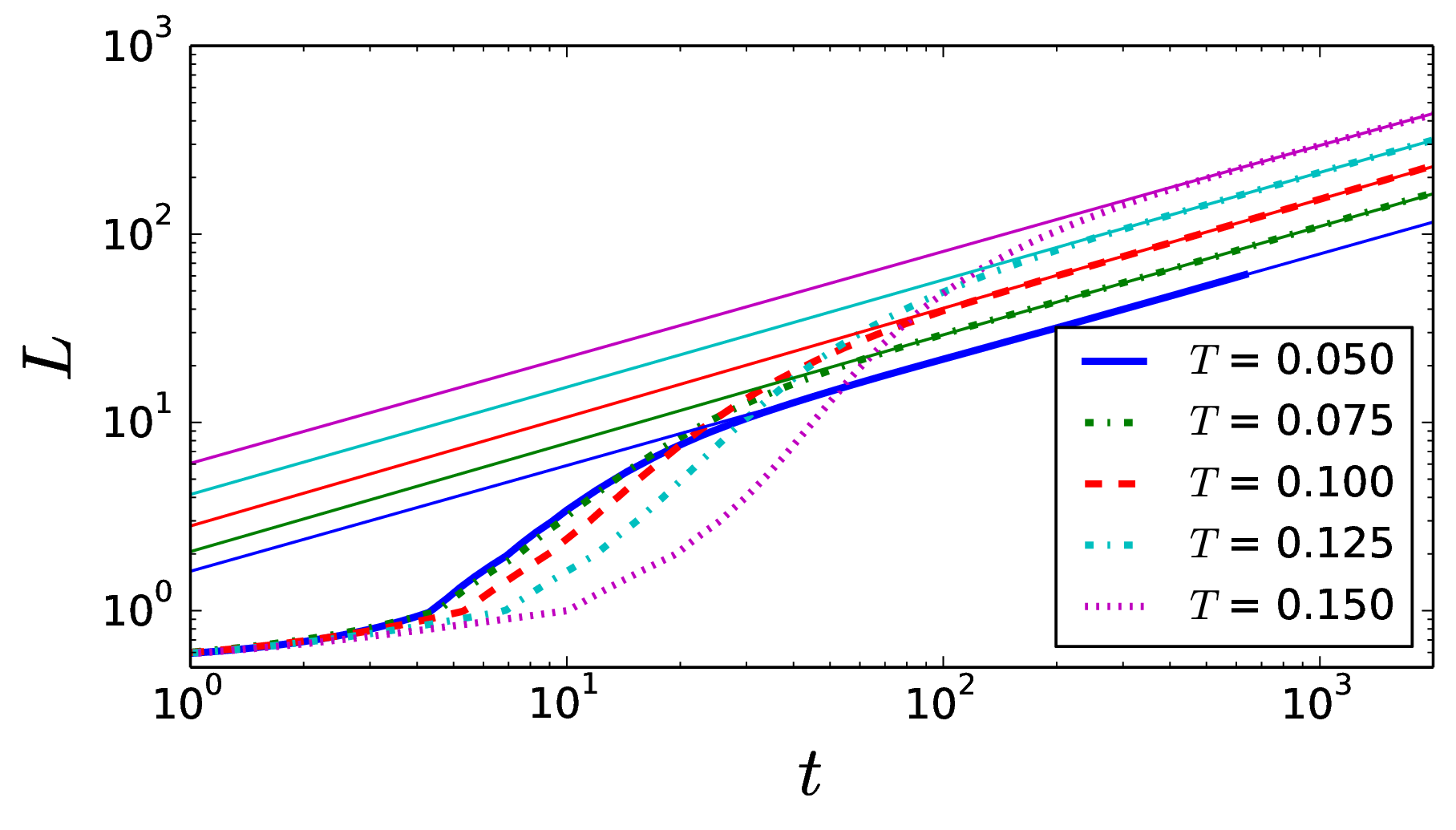}
  \caption{\label{f:LL}Thick lines are the scaling factor $L(t)$ for simulation with Langevin dynamics, averaged over $400$ runs, at different temperatures. Fixed parameters of simulations are  $\sigma=0.6$, $N=65536$ and $\gamma=1$. Straight thin lines are the function $c \, t^{1/z}$ with fitted $1/z$ and $c$.} 
\end{figure}

\begin{figure}
  \includegraphics[width=\linewidth]{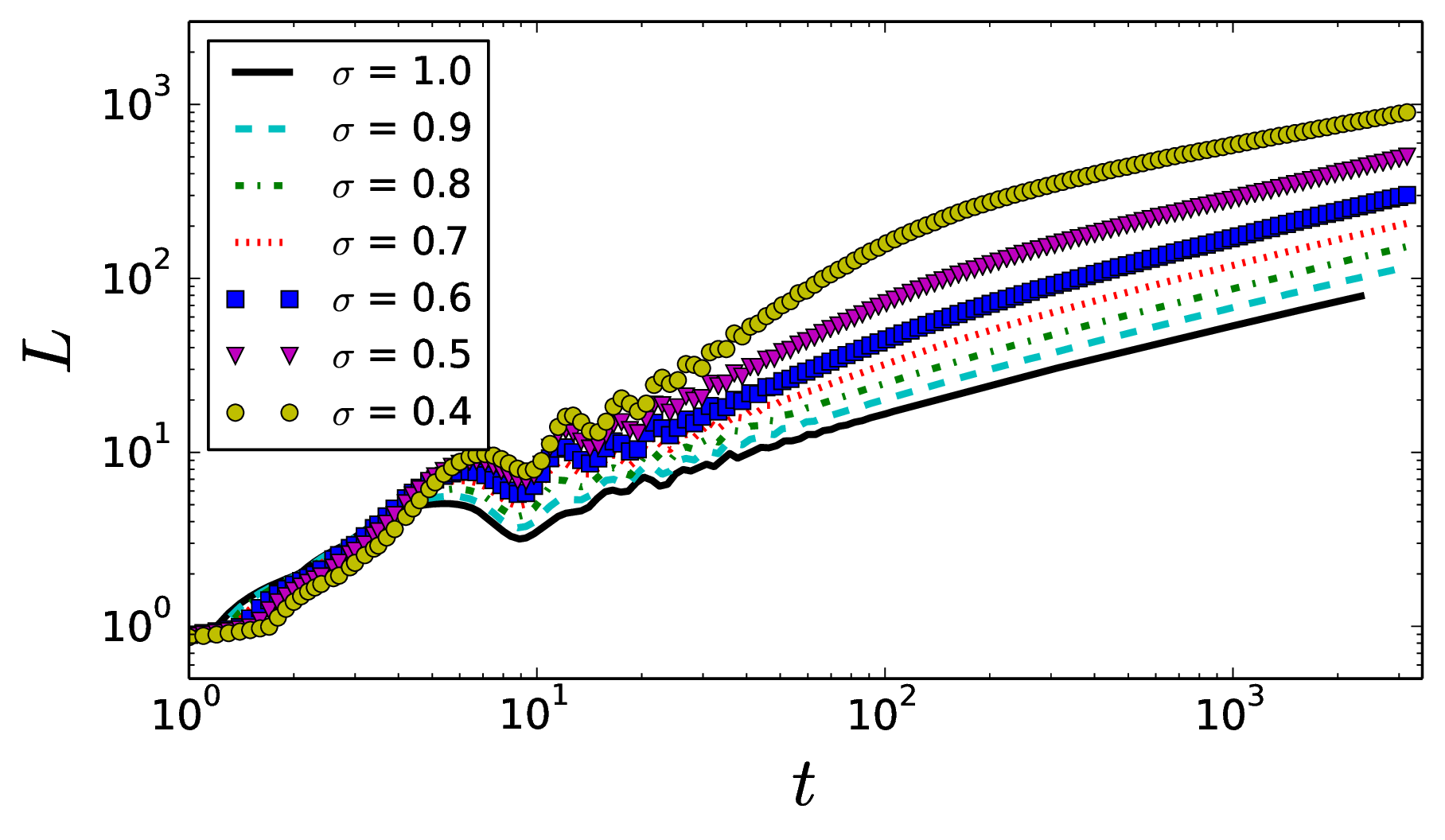}
  \caption{\label{f:LH_sigma}Scaling factor $L(t)$ for the Hamiltonian dynamics for different $\sigma$, averaged over $400$ runs. Fixed parameters $N=65536$ and $e = -0.06$.}
\end{figure}

\section{Langevin dynamics}\label{s:Ld}
\begin{figure}
  \centering\includegraphics[width=\linewidth]{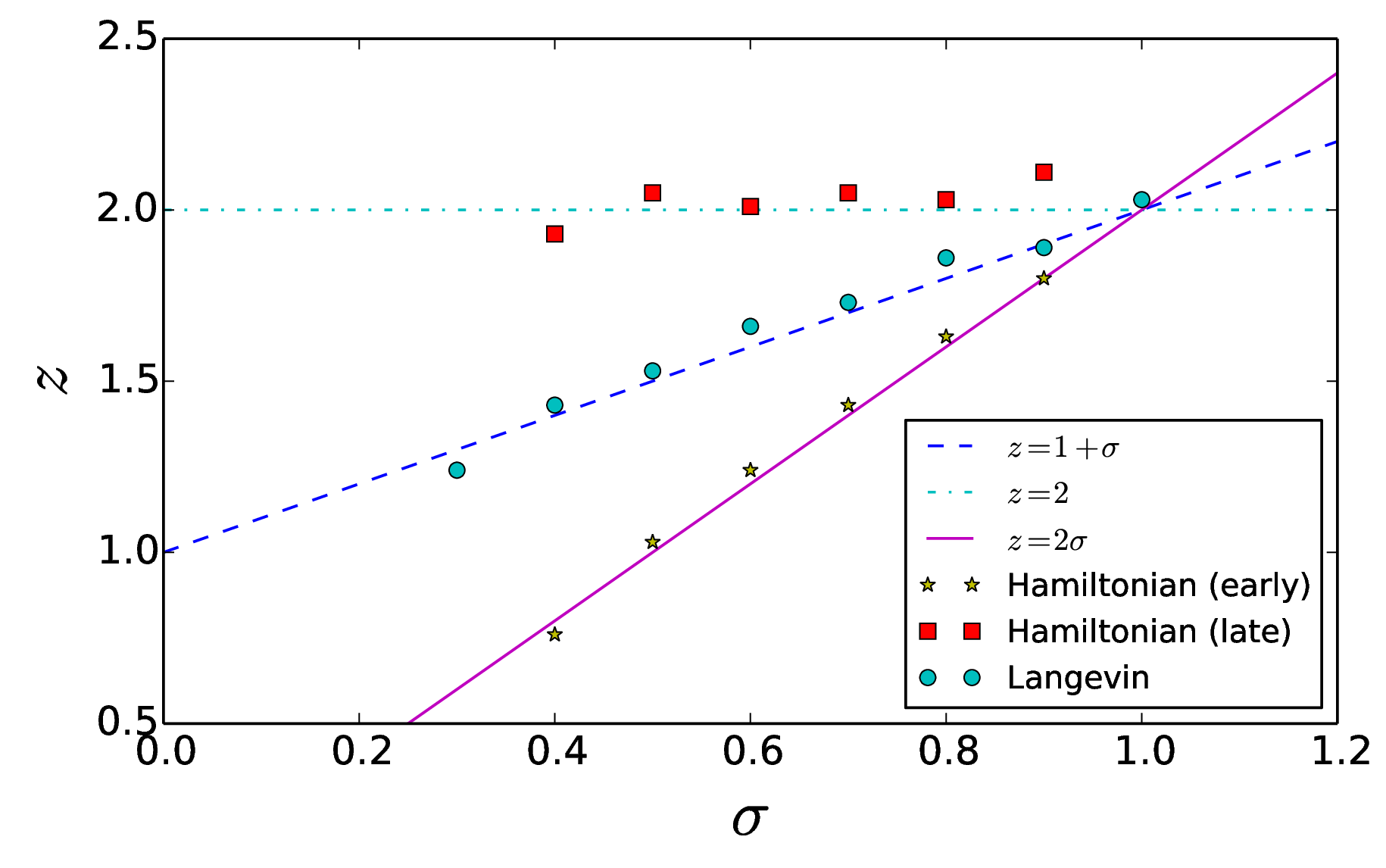}
  \caption{\label{f:main} $z(\sigma)$ as obtained from Langevin and Hamiltonian dynamics for simulations realized with $N=65536$ particles, averaged over $400$ realizations and with $T= 0.05$ and $\gamma = 1.0$ for Langevin dynamics, and $e =-0.06$ for Hamiltonian dynamics.} 
\end{figure}

For the Langevin dynamics, we chose $n =2$. After a transient time, the scaling factor $L$ grows as a power of time, and~\fig{f:LL} shows $L(t)$ for different values of temperature. One observes in these curves that there exists a transient regime $t<t_{\mathrm{trans}}$, where $t_{\mathrm{trans}}$ depends on temperature, followed by a regime where the slope does not depend on temperature. The curves after the transient time can be empirically fitted by the function:
\begin{equation}\label{e:guess}
 L(t_0) + c \, (t-t_0)^{1/z}\,, \quad t>t_0,
\end{equation}
using $1/z$, $c$ and $t_0$ as fitting parameters, and where the numerical curve $L(t_0)$ has been interpolated to have a smooth form for it.
Determined by this procedure, $z(\sigma)$ is in good agreement with
\begin{equation}\label{e:zLang}
  z_{c} (\sigma) = 1 + \sigma\,,
\end{equation}
 which corresponds to the law predicted for the defect dynamics \cite{Cardy1993,Bray1994b} (see~\fig{f:main}). 

If instead of~\eqref{e:guess}, we use for the fit the simple power law $c \, t^{1/z}$, one can find a systematic shift of the exponent $z$ fitted of roughly $+0.2$. This fit is valid for a smaller temporal window, but works also at larger values of the temperature (e.g. $T \sim 0.15$) where~\eqref{e:guess} presents some convergence issues.
For the temperatures where the fit~\eqref{e:guess} works, varying $\gamma$ or $T$ affects the transient regime, i.e. parameters $t_0$ and $c$, but no significant effect on $z$ was found, and the same holds at every temperature using: $ c \, t^{1/z}$. This can be clearly seen also in \fig{f:LL} where the slopes of the curves after the transient are all very close.

\begin{figure}
  \includegraphics[width=\linewidth]{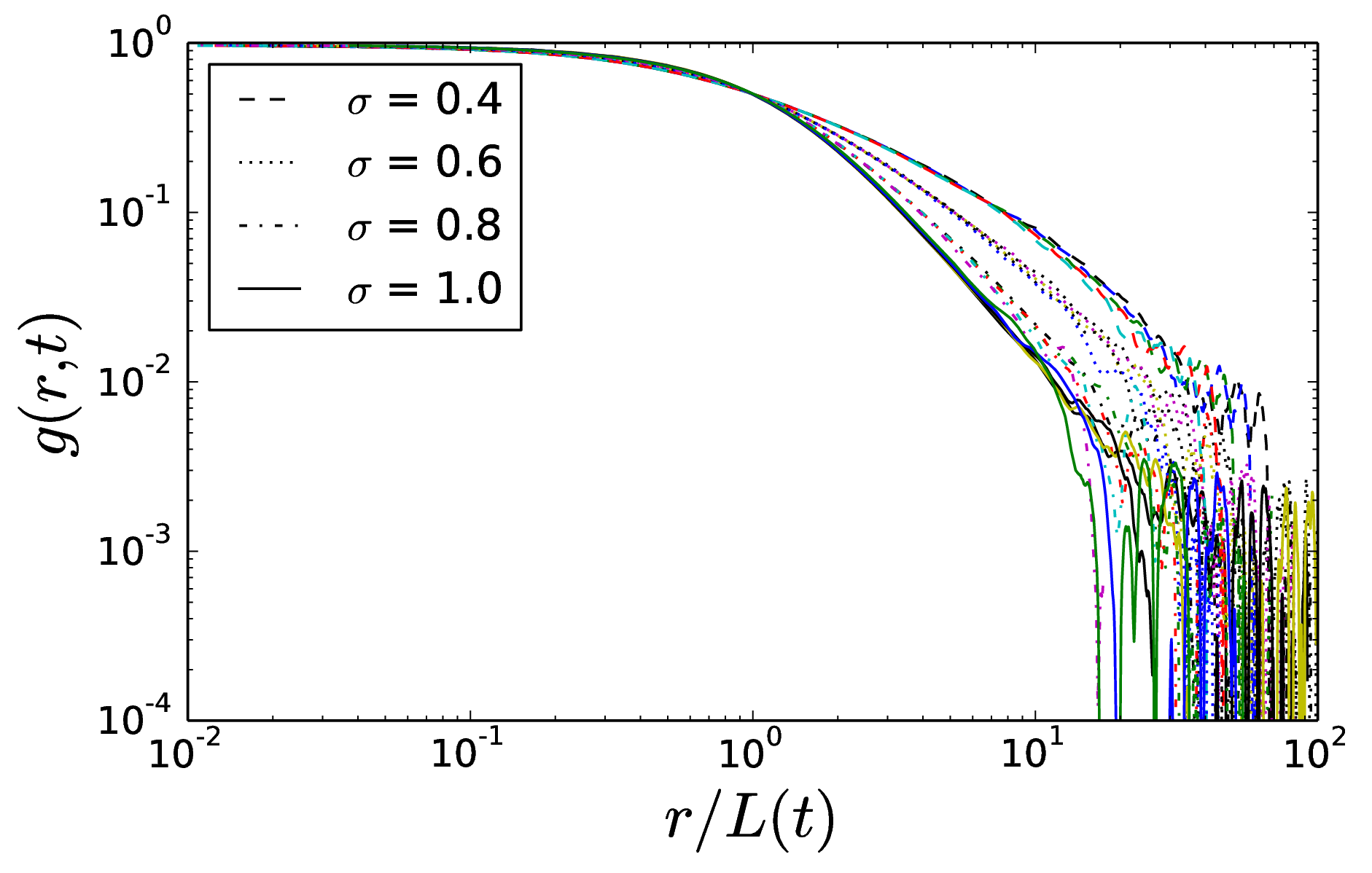}
    \includegraphics[width=\linewidth]{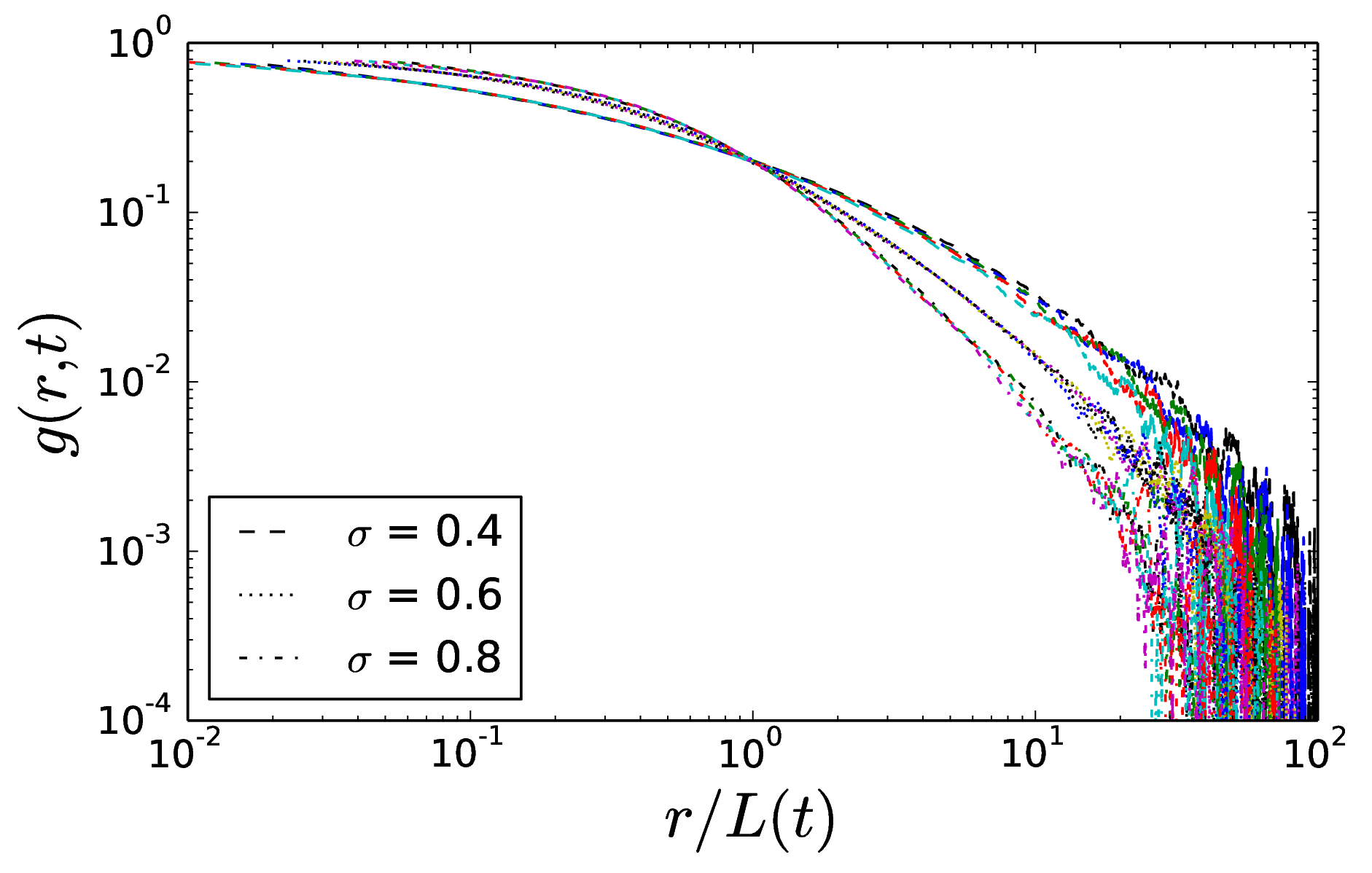}
    \includegraphics[width=\linewidth]{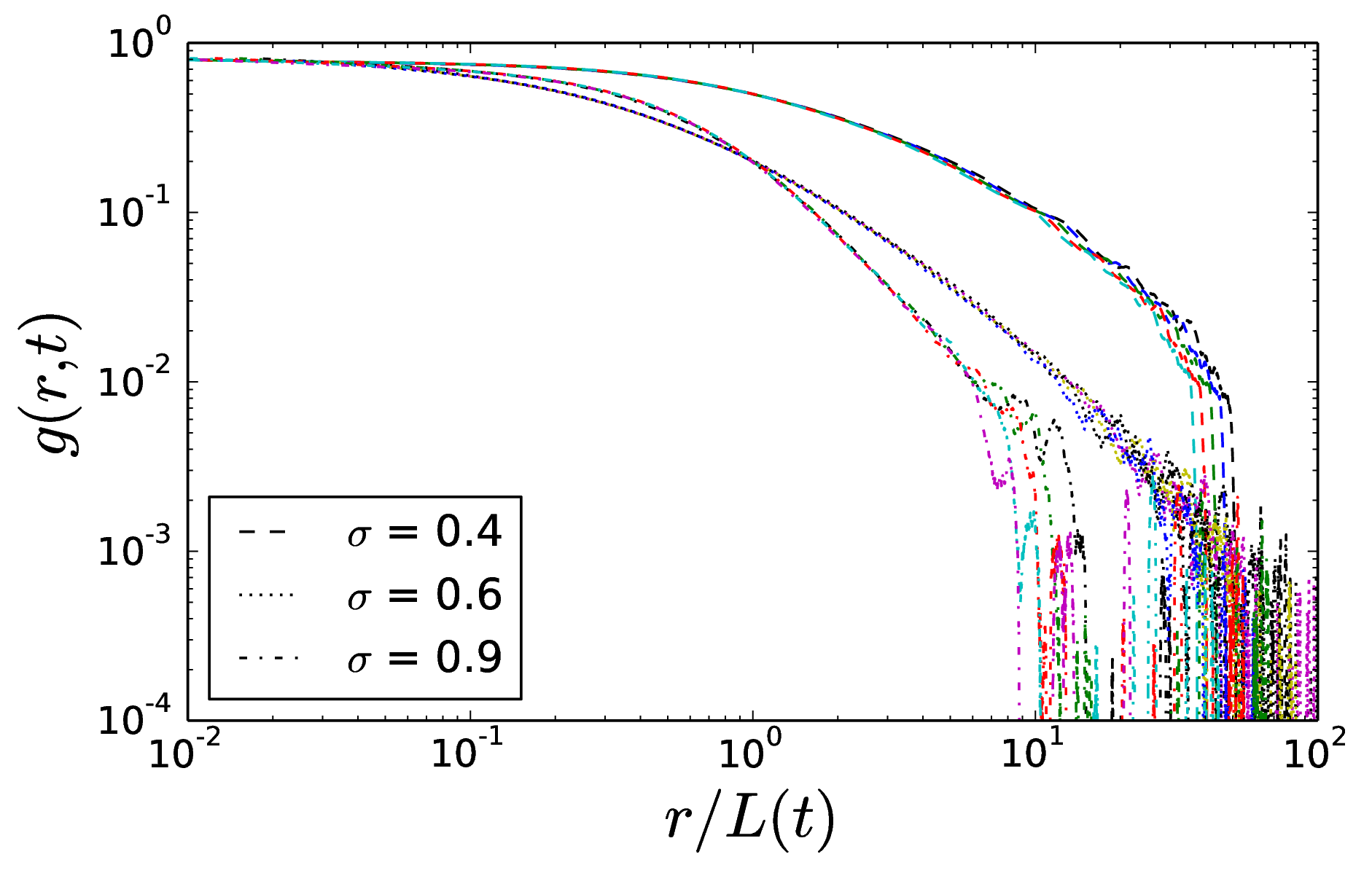}
  \caption{\label{f:scaled}  Correlation functions plotted in units of scaled distance $\tilde{r}=r/L(t)$, for five different times in the scaling regime, for each of the different values of $\sigma$ and the same other parameters as Fig.~\ref{f:g}. The top plot corresponds to the Langevin case, the middle plot to the Hamiltonian case at early times, and the lower to the Hamiltonian case at late times. $L(t)$ has been defined in~\eqref{e:guess} for the top plot and as $L(t)= c\, t^{\frac{1}{z}}$ in the others, and the values of $z$ are the same as in Fig.~\ref{f:main} for each value of $\sigma$.}
\end{figure}

Using the parameters of~\eqref{e:guess} obtained from the fits, we are able to confirm the validity of the scaling hypothesis for the Langevin dynamics. Indeed as can be seen in Fig.~\ref{f:scaled}, where $\tilde{g}(\tilde{r})$ is plotted, the correlation functions collapse very well.

\begin{figure}[h!]
    \includegraphics[width=\linewidth]{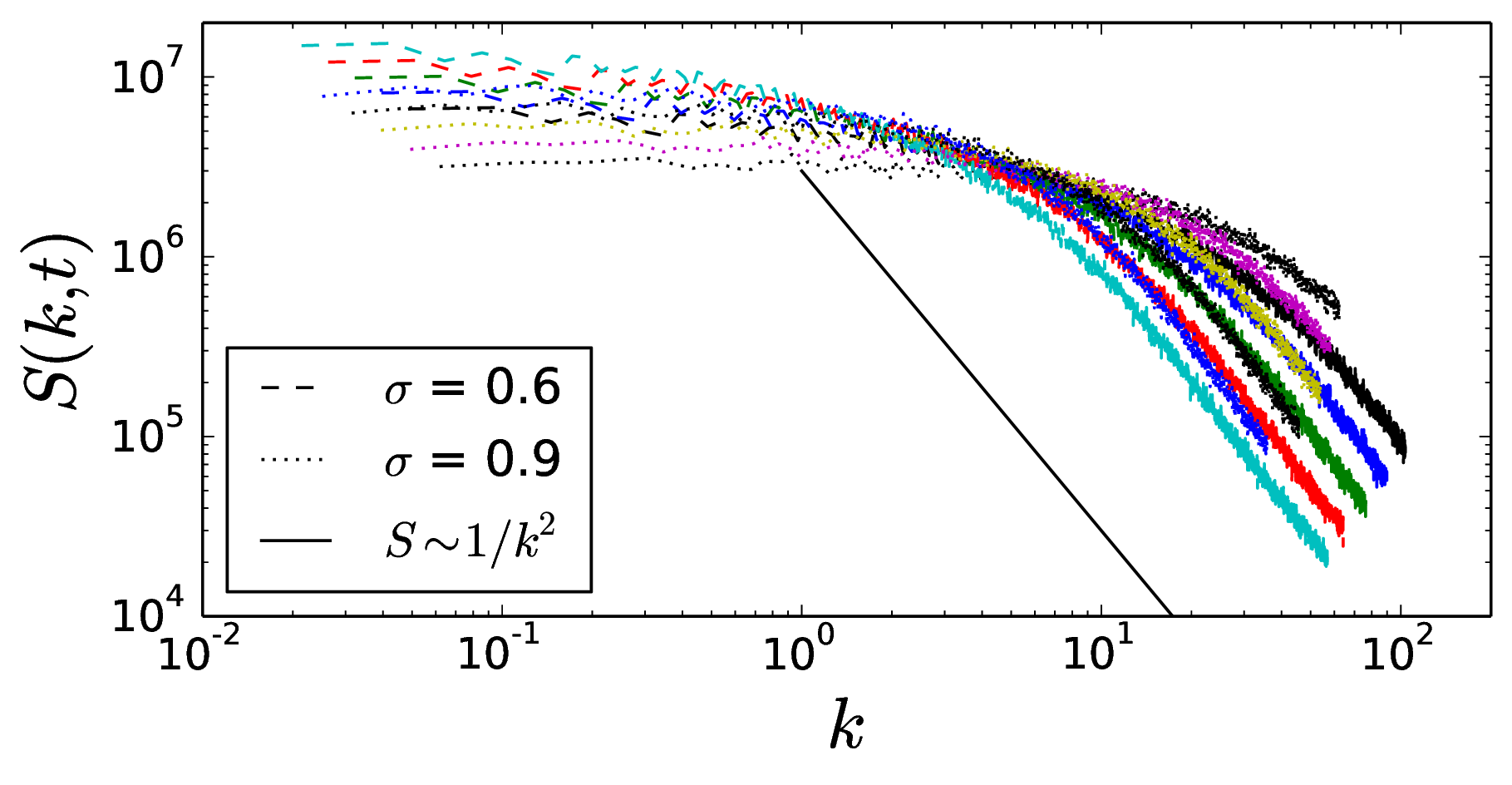}
    \includegraphics[width=\linewidth]{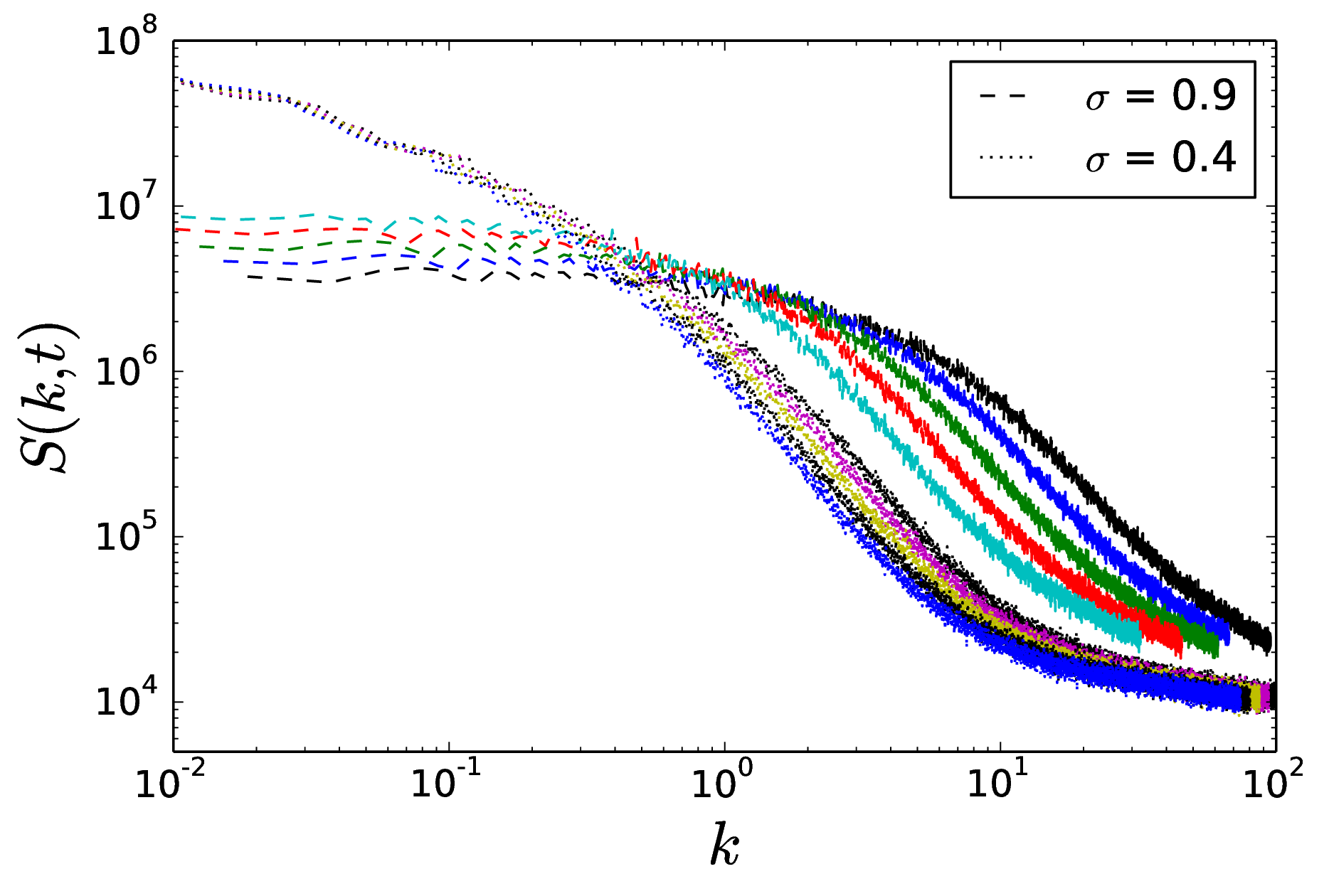}
  \caption{\label{f:strfac} Structure factor for Langevin (Top), and Hamiltonian (Bottom) simulations for different $\sigma$ and the same other parameters as Fig.~\ref{f:g}.
At large $k$ the Langevin case follows Porod's law~\eqref{eq:porod}, while the Hamiltonian shows a deviation.}
\end{figure}

Finally we analyse the domain structure: Porod's law~\cite{Porod51,Porod52,Furukawa78} relates the structure of defects in the order parameter, which are the sharp domain boundaries, to the tail of the structure factor $S$. Since the shape of defects affects the small distance behaviour of the correlation function, it can be analysed by looking at the large wave-vectors $k$ behaviour of the structure factor, which corresponds to its spatial Fourier transform. In a one dimensional case, the law predicts, for $k L \gg 1$:
\begin{equation}\label{eq:porod}
 S(k,t) \sim \frac{1}{L \, k^2 } \, \mbox{.}
\end{equation}
We confirm this law by plotting the structure factor in \fig{f:strfac}. This provides a further justification to the approach of~\cite{Cardy1993,Bray1994b}, which is based on the sharp domain boundaries approximation.

\section{Hamiltonian dynamics}\label{s:Hd}
Also in this case simulations exhibit scaling properties, though the behaviour is richer. We can notice in \fig{f:LH_sigma} the existence of an early
regime of power law growth of $L(t)$ which starts sooner than in the Langevin case, regardless of the energy of the system, where one does not need to exclude a transient time window to perform the fit. We can use $c \, t^{1/z}$ with only $1/z$ and $c$ as fitting parameters.
Here the best fit is given choosing $n = 5$, and we used it to define $L(t)$ in the microcanonical ensemble. The values of the dynamic exponent, plotted in Fig.~\ref{f:main}, stay very close to the line:
\begin{equation}\label{e:zHam_early}
  z_{\mu}^{early}(\sigma) =2 \sigma\, .
\end{equation}
At larger times simulations show that the factor $L(t)$ starts deviating from a straight line and the system displays a crossover towards another scaling regime ($t > 100$ in Fig.~\ref{f:LH_sigma}). The dynamic exponent, shown in 
Fig.~\ref{f:main}, now stays close to the law:
\begin{equation}\label{e:zHam_late}
  z_{\mu}^{late}(\sigma) = 2 .
\end{equation}
The collapse of the correlation functions has been checked, for both regimes, in Fig.~\ref{f:scaled}.
These relations constitute an original result pertaining to Hamiltonian coarsening dynamics and differentiate it from the Langevin dynamics. 
For the peculiar case $\sigma = 1.0$, there are several time windows which provide an equally good collapse but different exponents ($\pm 0.3$), whereas for other values of $\sigma$ exists a time window that gives an optimal fit and an optimal collapse of the correlation function.
This can be interpreted by the fact that for $\sigma =1.0$, as in~\cite{Nam12}, a logarithmic correction to the power law may be needed. Using a similar ansatz: $L(t) \sim (t \, \log t)^{1/z}$ we can identify an optimal time window for the fit and collapse, which results in $z = 2.42$, although other forms of $L(t)$ give equally good fits, which makes hard to give conclusive statements about this case.

In the limit $\sigma \to 0$, we are not able to extract any dynamical exponent for Hamiltonian dynamics since the system relaxes quickly to a macroscopically magnetized phase. In the context of gravitation, this short transient was termed ``violent relaxation''~\cite{LyndenBell67} and does not depend significantly on the system size. This property was later shown to be common to many Hamiltonian systems with long-range interactions and it appears, starting from a thermal state, also after a quench~\cite{Gupta_2016}.
We have indeed observed that the early regime of the Hamiltonian simulations displays additional oscillations (\fig{f:LH_sigma}). They persist in the thermodynamic limit, which shows that they are a result of collective oscillations. Such oscillations are typical of the dynamics of inertial systems with the mean-field potential ($-1<\sigma<0$)~\cite{BSDdNR2011} and therefore we can safely conclude that some of these effects persist into the regime of the coarsening dynamics. The fact that we do not see them for Langevin dynamics, where inertia is not present, suggests that they may be a feature of Hamiltonian dynamics.
Finally the analysis of the structure factor shows another difference with respect to the Langevin case: as can  be seen in \fig{f:strfac} the Porod's law is not satisfied for Hamiltonian dynamics since there is no power-law decay at large $k$.

\section{Discussion}

Effective models of energy conserving systems, obtained through a coarse-graining of the microscopic model, are characterized by a temperature (or energy) field coupled to the order parameter field~\cite{Halperin74,Umantsev88,Fried93,Truskinovsky93,Penrose1994}. These models can be considered good approximations of our Hamiltonian case,  and the temperature field can be considered, assuming that the system is in a quasi-equilibrium state, as equivalent to the mean kinetic energy in a small spatial region. For the Langevin case, which does not have a kinetic term, coarse-grained models are instead characterized by a single field which represents the order parameter. In these models the law $z_c = 1 +\sigma$ can be understood considering sharp domain boundaries in the order parameter field, which evolve driven by an effective interaction, as in~\cite{Cardy1993,Bray1994b}.
To understand our results for the microcanonical $\phi^4$ model we have to look qualitatively at the evolution of the magnetization and of the temperature.   
When a magnetization domain disappears, the potential energy of the system decreases because of the disappearance of its two boundaries. Since the total energy is conserved, the potential energy lost is transformed, locally, into kinetic energy which will diffuse by thermal conduction. 
This creates inhomogeneity in the temperature field which makes the relaxation  of systems easier in the areas of larger temperature. 
This means that the temperature dynamics can thus drive the domain walls, and the fact that the temperature evolves diffusively justifies the diffusive relaxation of the order parameter $z = 2$.
This mechanism has been illustrated in the case of an energy conserving coarse-grained model with nearest-neighbour interactions for a single domain interface in~\cite{Umantsev02}.

In this letter, we provide evidence that, for a one-dimensional $\phi^4$ model with algebraically decaying interactions, Hamiltonian and Langevin dynamics generate coarsening regimes in which the scaling hypothesis~\eqref{e:hyp} is valid, but the laws for the dynamical exponents $z(\sigma)$ appear to be different.  For the Langevin dynamics our results show, when using a formula that allows to exclude the transient regime, that the dynamic exponent depends on the exponent of the interaction potential $\sigma$ according to the law: $z_{c} = 1 + \sigma$, in full agreement with the law obtained from the dynamics of sharp interfaces.
For Hamiltonian dynamics, we show that this approximation is not valid since the temperature diffusion, which does not appear in the Langevin case, is coupled to the order parameter profile.  In this case, we find  at early times the new empirical law: $z_{\mu}^{early} = 2 \sigma$, and in the asymptotic regime, in which the relaxation of the order parameter is dominated by the diffusive dynamics of the temperature field, the law: $z_{\mu}^{late} = 2$. 
These results show that the effect of the contact with the environment is crucial for coarsening dynamics.

\section{Acknowledgements}

We thank R. Pa\v{s}kauskas and S. Ruffo for their involvement in this the project and their essential suggestions.

\bibliographystyle{eplbib}
\bibliography{Coarsening_a}

\end{document}